\documentclass[]{aastex631}

\usepackage{natbib}
\bibpunct{(}{)}{;}{a}{}{,}

\usepackage{xspace}

\newcommand{\Swift}{\textsl{Swift}\xspace}

\newcommand{\Fermi}{\textsl{Fermi}\xspace}

\newcommand{\nus}{\textsl{NuSTAR}\xspace}
\newcommand{\RXTE}{\textsl{RXTE}\xspace}
\newcommand{\gm}{$\gamma$}

\newcommand{\pks}{PKS\,2005$-$489\xspace}
\usepackage{CJK}

\received{09 Jan, 2023}
\revised{07 Mar, 2023}
\accepted{16 Mar, 2023}
\submitjournal{ApJ}

\shorttitle{The peculiar X-ray spectrum of 2005-489}
\shortauthors{Chase et al.}

\graphicspath{{./}{figures/}}
\usepackage{rotating}

\begin{document}
\begin{CJK*}{UTF8}{gbsn}

\title{The peculiar variable X-ray spectrum of the active galactic nucleus PKS\,2005$-$489}


\author[0000-0002-0304-5701]{Owen Chase}
\affiliation{Department of Astronomy \& Astrophysics, The Pennsylvania State University, University Park, PA 16802, USA}

\author[0000-0001-6191-1244]{Felicia McBride}
\affiliation{Department of Physics and Astronomy, Bowdoin College, Brunswick, Maine 04011, USA}

\author[0000-0002-5726-5216]{Andrea Gokus}
\affiliation{Department of Physics, Washington University in St. Louis, One Brookings Drive, St. Louis, MO 63130, USA}
\affiliation{McDonnell Center for the Space Sciences, Washington University in St. Louis, One Brookings Drive, St. Louis, MO 63130, USA}
\affiliation{Dr. Karl Remeis-observatory \& ECAP, Friedrich-Alexander Universit\"at Erlangen-N\"urnberg, Sternwartstr. 7, 96049 Bamberg, Germany}
\affiliation{Lehrstuhl f\"ur Astronomie, Universit\"at W\"urzburg, Emil-Fischer-Strasse 31, 97074 W\"urzburg, Germany}

\author[0000-0002-2235-3347]{Matteo Lucchini}
\affiliation{MIT Kavli Institute for Astrophysics and Space Research, Cambridge, MA 02139, USA}

\author[0000-0001-9826-1759]{Haocheng Zhang}
\affiliation{NASA Postdoctoral Program Fellow}
\affiliation{NASA Goddard Space Flight Center
Greenbelt, MD 20771, USA}

\author{Roopesh Ojha}
\affiliation{NASA HQ, 300 E St SW, Washington DC, 20546-0002, USA}

\author[0000-0002-3714-672X]{Derek B. Fox}
\affiliation{Department of Astronomy \& Astrophysics, The Pennsylvania State University, University Park, PA 16802, USA}

\begin{abstract}
\pks is a well-known, bright southern BL Lac object that has been detected up to TeV energies.
In a low-flux state it exhibits the expected multiwavelength double-peaked spectrum in the radio -- $\gamma$-ray band. The high-flux state shows extreme flux variations in the X-ray band with a hardening as well as a peculiar curved feature in the spectrum. Thus far, \pks is the only source to exhibit such a feature.
To study the X-ray variability further, we obtained the first hard X-ray spectrum of the source with \nus. We compare quasi-simultaneous radio, optical, UV, soft and hard X-ray, and $\gamma$-ray data of \pks to archival data in order to study its broadband behavior. We find a very consistent quiet state in the SED, with little variation in spectral shape or flux between the 2012 and 2020 data.
A possible explanation for the peculiar X-ray spectrum in the flaring state is an additional component in the jet, possibly accelerated via magnetic reconnection, that is not co-spatial to the low-flux state emission region. 
\end{abstract}


\section{Introduction} 
\label{sec:intro}
Active galactic nuclei (AGN) are among the most luminous persistent sources we observe.
They can be divided into radio-quiet and radio-loud sources based on the absence or presence of a relativistic outflow of plasma, called a jet. A radio-loud AGN whose jet is oriented at small angles with respect to our line of sight is called a ``blazar" \citep{blandford1978} and often exhibits rapid variability at all wavelengths. 
Blazars can further be subdivided into two classes, flat-spectrum radio quasars (FSRQs) and BL Lacertae (BL Lac) objects \citep{urry1995}. The two classes likely differ in jet power, accretion rate and magnetic field strength. Observationally, FSRQs are identified by their strong emission lines, which are weak or absent in BL Lac type objects.
BL Lac type objects (at weaker jet power) are typically observed as high-energy peaked objects --- i.e., their spectral energy distributions (SEDs) show synchrotron and high-energy peaks that are observed at $\gtrsim 10^{15}$~Hz and $\gtrsim 10^{23}$~Hz, respectively.
This makes these sources ideal targets for Very-High Energy (VHE) observatories, including H.E.S.S., MAGIC, VERITAS, FACT, and the Cherenkov Telescope Array (CTA).
In fact, of the 279 sources listed in TeVCat, 84 are AGN, including 57 high-energy peaked blazars (HBLs) or BL Lac type sources, one of which is \pks \citep{tevcat}.

\pks is a BL Lac type blazar at coordinates $\alpha=302\fdg 3558, \delta=-48\fdg 8316$ \citep{gaia_dr2}, and redshift $z=0.07067$ \citep{keeny2018}. Its host galaxy is a giant elliptical galaxy \citep{Scarpa2000}.
\pks was detected by Parkes in the 2.7\,GHz survey in 1972 \citep{parkes}. EGRET (20 MeV -- 30 GeV band) was the first telescope to make a $\gamma$-ray detection of the source \citep{egret}, which was later confirmed in the 30 GeV -- 100 TeV VHE band by the H.E.S.S. telescopes in Namibia \citep{hess10} .
The extreme X-ray variability of the source was first seen in EXOSAT data by \citet{Giommi1990}, who describe flux changes of a factor of 35$\pm$7 in the 0.7--8\,keV band, as well as variability on 4--5 hour timescales and a hardening of the spectrum. 
\citet{Sambruna1995} study the broadband behavior of \pks and find similar variability in \textsl{ROSAT} observations.
The X-ray spectrum is analyzed in more detail by \citet{Perlman1999} using \textsl{RXTE} data. 
They report on a flare that occurred in 1998, during which the 2--10\,keV flux of the source increased to $3.33\times10^{-10}$\,erg\,cm$^{-2}$\,s$^{-1}$, an increase in flux by a factor of 30.
\citet{Perlman1999} discuss the origin of the flare as particle injection into the jet or \textsl{in situ} particle acceleration with a hardened X-ray spectrum resulting from synchrotron cooling.
\textsl{BeppoSAX} observations taken during the 1998 flare reveal a slightly curved spectrum \citep{Tagliaferri2001}.

X-ray observations with \textsl{XMM-Newton} and \textsl{RXTE} in 2009 revealed an unexpected  spectral break at 4\,keV. Strong X-ray variability was observed (flux increased by a factor of $\sim 50$), while the TeV $\gamma$-rays do not show strong variability (flux increase by a factor of $\sim 2$). A possible explanation is electron scattering during high-flux states occurring primarily in the Klein-Nishina regime, resulting in less efficient photon production at TeV energies.
The \citet{hess2011} provided multiwavelength data from the 2009 flare in which the flux decreased after the flare in both the \Fermi/LAT and optical bands. The SED was modeled with a one-zone SSC model, where the X-ray spectrum was explained as a break in the synchrotron emission. The authors also reported that the VHE $\gamma$-ray flux hardly changed from quiescence while the X-ray flux varied drastically \citep{hess2011}.

No other multiwavelength campaigns of this object has been published since 2011, excluding the work by \citep{krauss2016}, and no other BL Lac object show a comparable X-ray flaring behavior as far as we know.

\citet{krauss2016} presented quasi-simultaneous data for the source that included a period of low emission ($\sim$ 2012) and a period of high-emission (2009). 
The SED shows a deviation from the expected parabola shape, with a hardened X-ray spectrum in the high state that is incompatible with the expected synchrotron peak.
The origin of the peculiar X-ray feature remains unknown. 
It is possible that this is an additional jet component, an extreme peak shift of the synchrotron peak, an additional hadronic component (proton synchrotron or a pion cascade), or a signature of magnetic reconnection.
However, the question remains why this seems to be the only source to exhibit such a peculiar feature.

In this paper, we analyze newly obtained \textsl{Nuclear Spectroscopic Telescope Array (NuSTAR)} data for \pks and compare it with archival data in order to study the source's extreme variability in the X-ray band and to constrain the emission mechanisms at play during the quiescent and flaring states.
The paper is organized as follows. In Section 2, we discuss the observations and the methods used to analyze the multiwavelength data. In Section 3, we describe and discuss the results of our observations.
In Section 4, we present our conclusions.

Throughput the paper, we use a cosmological model with $\Omega_m=0.3$, $\Lambda=0.7$, $H_0=70$\,km\,s$^{-1}$Mpc$^{-1}$ \citep{Beringer2012}.

\section{Multiwavelength Observations} 
\label{sec:obs}

In this section, we describe the multiwavelength data we extracted and analyzed, including archival non-simultaneous data that were added to the SED.
We analyze data in three different time periods: 2008--2009 ($\alpha$), 2011--2012 ($\beta$), and 2020 ($\gamma$). Archival data was then added to the SED. The time ranges designated $\alpha$ and $\beta$ are identical to those defined in \citet{krauss2016}. The details of the epochs are listed in Table~\ref{tab-times}. Archival data refers to all other data sets included in the SED that are not quasi-simultaneous with the other three time ranges.

\begin{table}[bht]
    \centering
        \caption{Table of observation time ranges used throughout the paper. The 2020 ($\gamma$) time range is centered on the \nus time range, and the length depends on the instrument used. The $\alpha$ and $\beta$ time ranges are defined by \citet{krauss2016} by using a Bayesian Blocks analysis with \Fermi/LAT light curves.}
    \begin{tabular}{lccl}
         Label & Time [UTC] & Time [MJD] & time criteria  \\
         \hline
          $\alpha$ & 2008-08-04 -- 2009-08-03 & 54682.6552778 -- 55046.6552778 & \Fermi/LAT light curve\\
          $\beta$ & 2011-03-14 -- 2012-03-15& 55634.6552778 -- 56264.6552778   & \Fermi/LAT light curve \\
          $\gamma$ &  2020-11-28 07:06:09 & 59181.2959375  & \nus observation \\
    \end{tabular}
    \label{tab-times}
\end{table}
 
\subsection{\nus} 
\label{sec:nustar}
\nus is a hard X-ray telescope with focusing optics that is sensitive in the 3\,keV-79\,keV energy range \citep{nustar}. \nus observed \pks on 2020-11-28 07:16:57 UTC (59181.3034 MJD) for about 64\,ks (ID 60601016002).  Data sets from focal plane modules A and B are grouped and treated as one data set unless otherwise noted. 

The raw data were reprocessed using the \texttt{nupipeline} tool (version 0.4.8) in the HEASoft data analysis package (version 6.28). Spectra were extracted from the resulting event files using the \texttt{nuproducts} tool.
Focal plane module A source data were extracted from within a circle of radius 77.364$^{\prime\prime}$ centered on coordinates 302\fdg3605, $-$48\fdg8330. A background region was defined with a circle of radius 211.153$^{\prime\prime}$ centered on coordinates 302\fdg2699, $-$48\fdg9151.
Focal plane module B source data were extracted from within a circle of radius 86.224$^{\prime\prime}$ at coordinates 302\fdg3541, $-$48\fdg8326, while the circular background region was centered at coordinates 302\fdg2576, $-$48.9028 with a radius of 234.367$^{\prime\prime}$.
The final spectra were analyzed in the \textsl{Interactive Spectral Interpretation System} \citep[ISIS;][]{Houck2000}.
Data were binned to a minimum signal-to-noise ratio of 20. As the background dominates the data above 25\,keV, we only used data in the 3--25\,keV range.

\subsection{\Fermi/LAT}
The \Fermi satellite observes the sky at \gm-ray energies. One of the two instruments onboard is the Large Area Telescope (LAT), a pair-conversion telescope, which is sensitive to energies of 20\,MeV up to more than 300\,GeV \citep{fermi2009}. The observing strategy of \Fermi/LAT is to monitor the whole sky, which is achieved on average after two orbits (3.2\,h). As a result, continuous \gm-ray data have been gathered since the launch of the satellite in 2008.
The \Fermi/LAT analysis has been performed using the \texttt{ScienceTools} version 1.2.23 and \texttt{fermipy} version 0.20.0. The time range chosen for the analysis is 120 days, i.e., from 2020 September 29, 00:00 UTC to 2021 January 27, 00:00 UTC (MJD 59121 to MJD 59241), which is centered on the date of the \nus observation.
We extracted SOURCE class events for an energy range between 100\,MeV and 300\,GeV in a region of interest (ROI) of $10^{\circ}$ around the position of \pks reported in the Fourth \Fermi/LAT source catalog \citep[4FGL;][]{fermi_4fgl}. We included events that fulfilled the flags \texttt{DATA\_QUAL>0} and \texttt{LAT\_CONFIG==1} and excluded events with a zenith angle $z\geq90^{\circ}$ in order to dispose of \gm-rays produced by Earth-limb effects. For the instrument response, we use the post-launch function \texttt{P8R2\_SOURCE\_V2}.
The model we used to describe the ROI includes all known \gm-ray sources from the 4FGL within $15^{\circ}$, as well as the Galactic diffusion model \texttt{gll\_iem\_v07} and the isotropic diffusion emission model \texttt{iso\_P8R3\_SOURCE\_V2\_v1}. 
For our source of interest, \pks, we choose a log-parabola model for the spectrum as defined in the 4FGL. 
The model parameters are optimised via a maximum likelihood analysis. The significance of the modeled \gm-ray signal is then determined via the test statistic $\mathrm{TS}=2\Delta\log(\mathcal{L})$, with $\mathcal{L}$ being the likelihood function representing the difference between two models --- one with a point source at the given source coordinates and one without\citep{mattox1996}.
We kept all spectral parameters free for \pks and all sources within $3^{\circ}$. We allow the normalization to be free for sources within $5^{\circ}$ and for those with a TS value higher than 500. In addition, the parameters in the models for isotropic and Galactic diffusion are kept free. Other sources are modelled with the spectral parameters reported in the 4FGL.
The TS value of \pks in the chosen time range is 170, which is $\sim13\sigma$, and we determine the flux to be $1.2\pm0.3\times10^{-8}\,\mathrm{ph}\,\mathrm{cm}^{-2}\,\mathrm{s}^{-1}$.
The best-fit results for the spectral parameters are $\alpha=1.44\pm0.17$ and $\beta=1.5\pm0.8$.

\subsection{\Swift}
The \textsl{Neil Gehrels Swift Observatory} is a multiwavelength mission with three instruments: the X-ray Telescope (XRT), the UltraViolet Optical Telescope (UVOT) and the Burst Alert Telescope (BAT). BAT, XRT, and UVOT are sensitive in the following wavelength ranges: 15--150\,keV, 0.2--10\,keV, and 170--600\,nm, respectively \citep{swift}. 

\Swift has observed \pks irregularly since its launch in 2004 through a TANAMI\footnote{\url{https://pulsar.sternwarte.uni-erlangen.de/tanami/}} fill-in program, additional Guest Observer (GO) programs, and Target of Opportunity (ToO) observations.
We used observations taken quasi-simultaneously with \nus, as well as archival data from the BAT 104-month catalog to compile the SED.
The XRT and UVOT data were reduced and extracted using HEASOFT
v.6.30.1\footnote{\url{https://heasarc.gsfc.nasa.gov/docs/software/heasoft/}}. 
BAT archival data has been taken from the 104-month catalog \citep{bat}.
\subsubsection{\Swift/XRT}
XRT data have been extracted using an annulus centered on R.A.$=302\fdg3558$, dec.$=-48\fdg8316$ with an inner radius of $11.787^{\prime\prime}$ and an outer radius of $47.146^{\prime\prime}$. The innermost region up to $11.787^{\prime\prime}$ has been excluded to correct for pile-up.
The background region has been extracted from an annulus surrounding the source with inner and outer radii of 1.886$^\prime$ and 2.947$^\prime$, respectively. We reduced and extracted the data using \texttt{xrtpipeline}, \texttt{XSELECT}, and \texttt{xrtmkarf}.
For spectral modeling, all X-ray data were binned with a minimum signal-to-noise ratio of 10. We used data in the 0.3--9\,keV energy band.

\subsubsection{\Swift/UVOT}
Data has been extracted using the \texttt{uvotimsum} and \texttt{uvot2pha} tools. The data were extracted using a $5^{\prime\prime}$ region centered on the source coordinates. 
We extracted the background from an annulus centered at the source coordinates with radii of 25$^{\prime\prime}$ and 50$^{\prime\prime}$, while ensuring that no other sources are present in the region.

\subsection{\RXTE/PCA}
The \textsl{Rossi X-Ray Timing Explorer (RXTE)} was an X-ray satellite designed to study the 2--200\,keV band up until its mission end in January 2012. We analyzed all available \RXTE data from its Proportional Counter Array (PCA), which is sensitive to the 3--60\,keV band. 
\RXTE observed \pks 164 times between 1997 and 2009. All spectra and backgrounds were re-analyzed and extracted using the standard tools in HEASOFT v6.28 provided through the publicly-available, python-based Chromos\footnote{\url{https://github.com/davidgardenier/chromos}} pipeline developed by \citet{Gardenier18}. This pipeline also provides a detailed description of the data reduction process. 
The spectra were extracted using only the standard-2 events in the proportional counter unit 2 (PCU2) as these data are best calibrated. We exclude data above 15\,keV due to an issue with the background modeling. Data were binned to a minimum signal-to-noise ratio of 15 for display purposes.
In Figs.~\ref{fig:full-sed}, \ref{fig:alpha-sed}, and \ref{fig:xspec}, we show the three different RXTE spectra. The highest flux state is observation 30432-01-03-00 (1998-11-10), the medium flux state is observation 94340-01-01-00 (2009-05-22; quasi-simultaneous with our $\alpha$ epoch), and the lowest flux state is 20342-03-01-02 (1997-07-30).

\subsection{ATCA}
The Australian Telescope Compact Array observes \pks regularly as a calibrator object, which is available through the ATCA calibrator database\footnote{\url{https://www.narrabri.atnf.csiro.au/calibrators/calibrator_database.html}}.
Flux measurements at 2.1, 5.5, 9.0, 17.0, 33.0, and 93.0\,GHz bands are available. Only the 93.0~GHz band lacks data that are contemporaneous to the \nus observation (2020-11-28). The data we chose for our analysis were taken between 2020-09-14 and 2021-02-21 and are considered quasi-simultaneous with the \nus observation.

\subsection{H.E.S.S.}
The High Energy Stereoscopic System (H.E.S.S.) telescopes are TeV Cherenkov $\gamma$-ray detectors located in Namibia. 
Archival data is available through the SSDC SED tool\footnote{\url{https://tools.ssdc.asi.it/SED/}}.
While the available data are generally not simultaneous to our observing periods, it is the only data set publicly available in the TeV band. One H.E.S.S. observation took place in 2009, which we consider quasi-simultaneous with the $\alpha$ state. We included H.E.S.S data in our modeling in order to constrain the SED at the highest energies \citep{hess10}.
As the data are not simultaneous, the model is potentially not accurate. However, when comparing the SED to the data used by \citet{krauss2016}, which lacks the TeV information, \Fermi/LAT data alone are not able to constrain the high-energy peak due to its position. Constraining the peak with the H.E.S.S. data is therefore preferable to having no constraint.

\subsection{Other multiwavelength data}
All remaining multiwavelength and archival data are identical to the data used by \citet[][]{krauss2016}, which includes information regarding the data reduction and analysis. Archival data (with the exception of the H.E.S.S. data) are not included in multiwavelength modeling of the SED.

\subsection{Multiwavelength data modeling}
\label{subsec:mwlmodel}
Details on the treatment of multiwavelength data are described by \citet[][Sections 2.4 and 2.5]{krauss2016}.
We treat multiwavelength data in a detector-space approach using ISIS v.1.6.2-48. We combine data with an assigned response (ARF/RMF; e.g. \Swift/XRT) with data sets that are only available as fluxes or flux densities. These data sets are assigned a diagonal matrix.
Unfolding of data is used for display purposes only; fluxes are calculated independently from the fit model \citep[for a detailed discussion, see][]{Nowak2005}.
Furthermore, we use a $\chi^2$ approach to modeling the SED and determining the best-fit. This is not statistically reliable because the data have been obtained by different instruments with 
different statistical backgrounds (e.g., likelihood analyses are required for \Fermi/LAT data), and the flux measurements obtained with the different instruments have not been cross calibrated. Therefore, $\chi^2$ values do not indicate an absolute model goodness-of-fit (i.e. interpreted as a probability); however, the relative reduced $\chi^2$ values between models of the same state and source are useful for finding a good empirical description.

Data in the X-ray, UV, optical, and infrared bands are affected by absorption and reddening effects. Photoelectric absorption is treated in the model by using the \texttt{tbabs} model \citep{Wilms2000} with the \texttt{vern} cross-sections \citep{Verner1996} and the \texttt{wilm} abundances \citep{Wilms2000}.
The $N_\mathrm{H}$ value for deabsorption and dereddening corrections ($3.63\times 10^{20}$\,cm$^{-2}$)  has been taken from the HI4PI survey \citep{hi4pi}.
We further apply the absorbing column to deredden the IR/optical/UV data by converting it to $A_\mathrm{V}$ from dust scattering halo measurements \citep{Predehl1995} with the corrections to the abundances of the ISM by \cite{Nowak2012}.

Systematic uncertainties are taken into account and added to the statistical uncertainties as described by \citet{krauss2016}.
For fitting the SED, we largely follow the procedure outlined in \citet[Section~2.5]{krauss2016}. We employ a double logarithmic parabola model that is characterized in \texttt{ISIS} as \texttt{(logpar(1)+logpar(2))*tbabs(1)*redden(1)}, which includes two parabolas, one for the synchrotron and one for the high-energy peak, as well as the absorption and reddening models. For the $\beta$ state, we include a blackbody model that indicates the presence of a weak host galaxy signature. This feature may be present in the 2020 $\gamma$ data, but due to the lack of optical information (with contemporary observations from only one UVOT filter), this cannot be confirmed. Therefore, we do not add a blackbody component to the model.
We reconstruct the SEDs first shown by \citet{krauss2016} with the 2009 and 2012 epochs, labeled $\alpha$ and $\beta$, respectively.
Then, we add our new 2020 data (dubbed $\gamma$) using the \nus and quasi-contemporaneous data discussed in Section~\ref{sec:obs}.
Previously, the source exhibited high and low fluxes during the $\alpha$ and $\beta$ states, respectively.
The following section describes the results from the modeling of the SED and X-ray data and discusses the implications.

\section{Results and Discussion}
\label{sec:results}
\subsection{The multiwavelength SED}
\begin{figure*}
    \centering
    \includegraphics[width=0.5\textwidth]{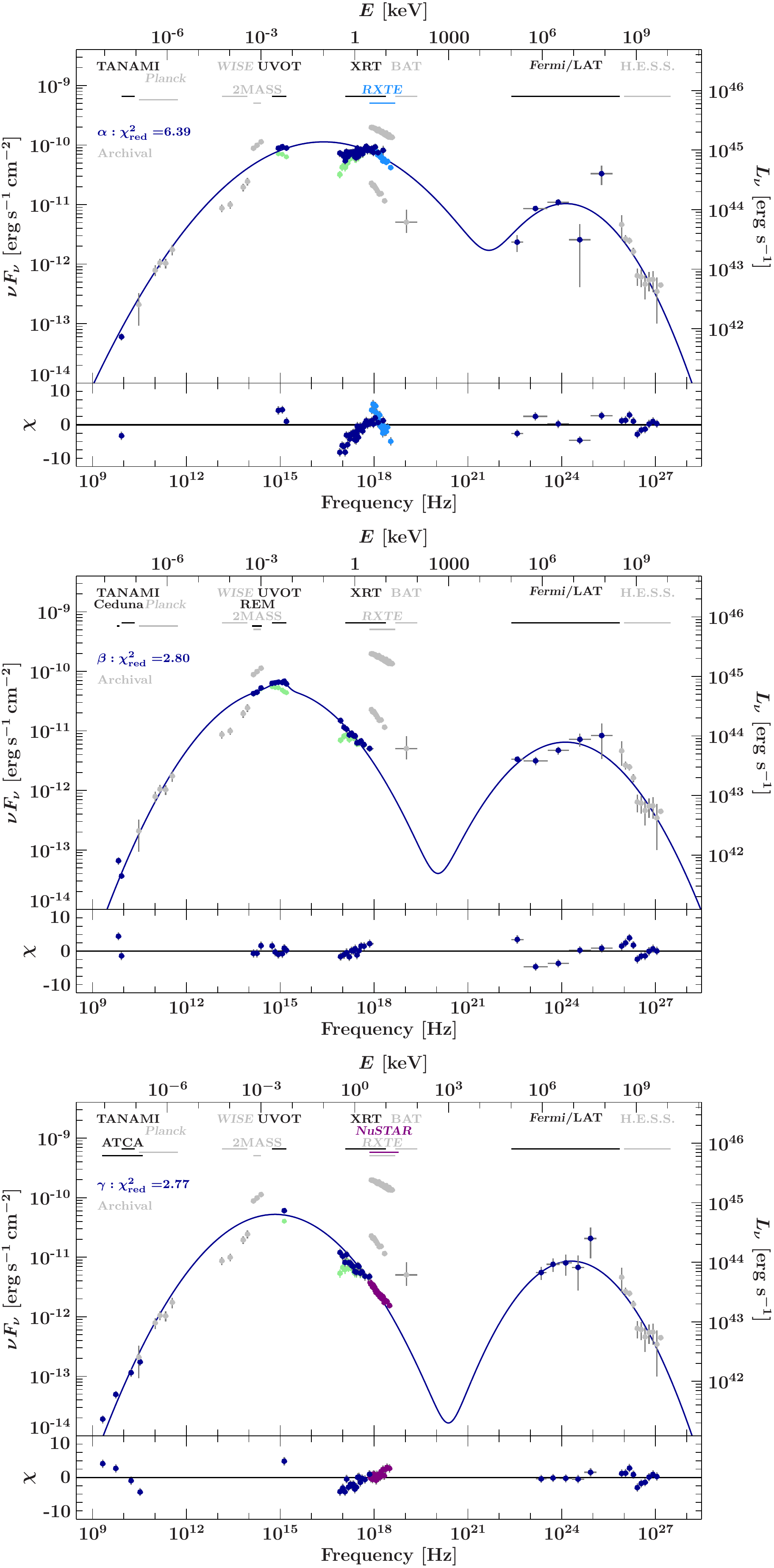}
    \includegraphics[width=0.49\textwidth]{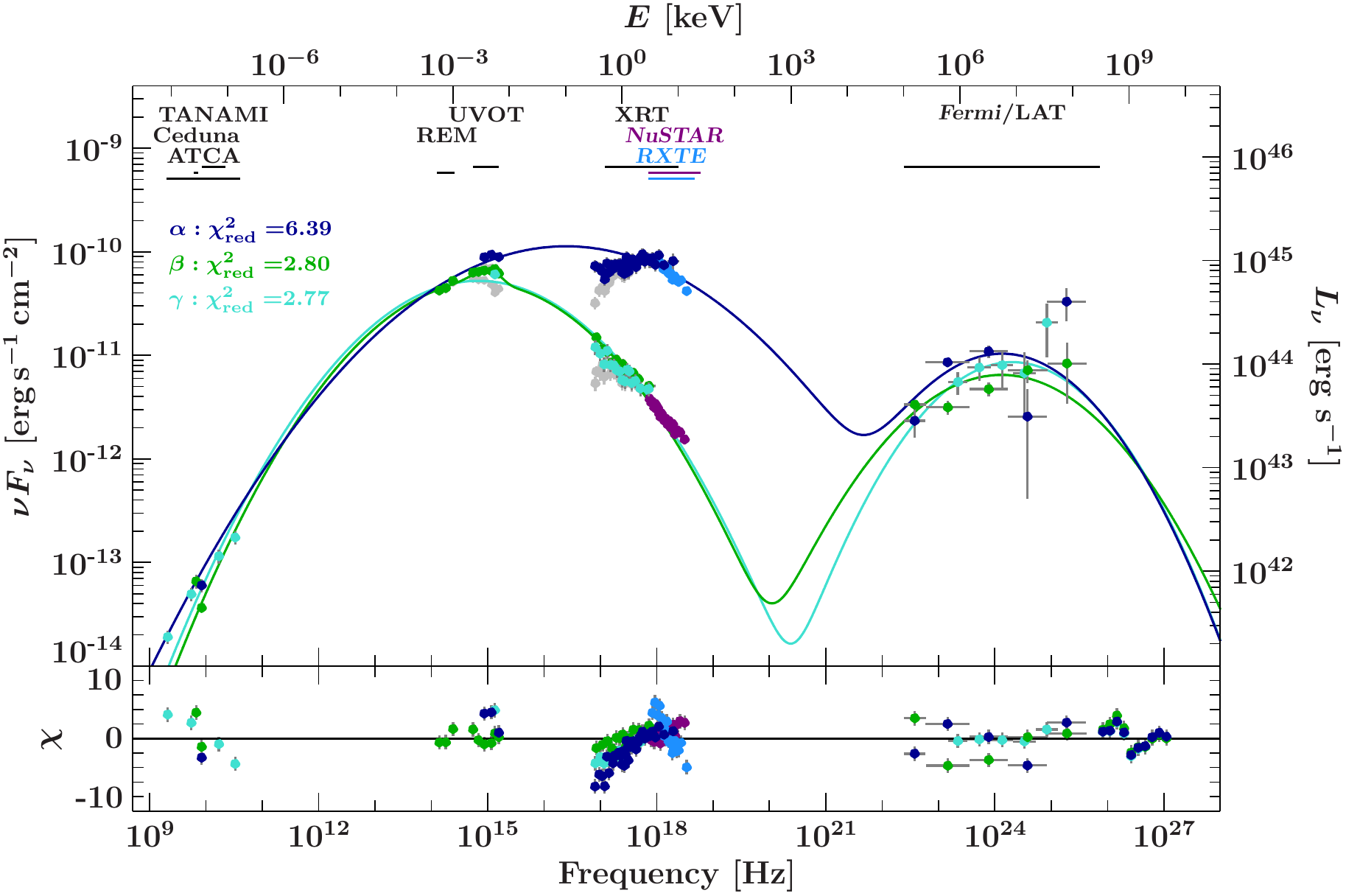}
    \caption{\textsl{Left:} The three panels show the three epochs $\alpha$ (2008--2009; top panel), $\beta$ (2011--2012; middle panel) and $\gamma$ (2020; bottom panel) in separate panels. Archival data (non-contemporaneous to the epoch shown) is displayed in gray. The instruments and their respective energy ranges are shown at the top of the plot, where gray indicates non-simultaneous data. In order to distinguish X-ray data, \nus data (simultaneous with the 2020 data, bottom left) is shown in purple. \RXTE data taken simultaneous with the $\alpha$ SED are shown in light blue (top panel). Observed IR, optical, UV, and X-ray data are shown in green, except when dereddened and deabsorbed, where they are shown in blue. Data used for fitting include the quasi-simultaneous data for each epoch (shown in color) and the archival H.E.S.S. data. Reduced $\chi^2$ values are shown in the top left side of the plot. The bottom panel shows the residuals for the fits to the parabolas.\\
    \textsl{Right:} The simultaneous data has been combined into one SED, where the three epochs have been separated into three different colors: $\alpha$ (2008--2009, blue), $\beta$ (2011--2012, green) and $\gamma$ (2020,turquoise).  The reddened and absorbed data is shown in gray. The data and models are identical to the panels presented on the left side, but are shown in one plot for easy comparison of fluxes and models.\\ 
    The only clear deviation from the model occurs in the $\alpha$ state in both the \Swift/XRT and \RXTE/PCA data, which exhibit a narrow curved feature incompatible with the broad SED peak expected from synchrotron emission.
    The quiescent state in the $\beta$ and $\gamma$ SEDs is virtually identical in flux.
    }
    \label{fig:full-sed}
\end{figure*}

We compiled the multiwavelength SED for the three epochs (2008--2009, $\alpha$; 2011--2012, $\beta$; and 2020, $\gamma$) with good quasi-simultaneous multiwavelength coverage (see Fig.~\ref{fig:full-sed}).
The \Swift data for $\alpha$ and $\beta$ have been reanalyzed with the updated calibrations and HEASoft tools, but the SED states consist of identical observations and data to those shown by \citet{krauss2016}.

In 2008--2009 ($\alpha$), the source was in a high-emission state with the X-ray flux reaching $\sim$20 times the flux of the quiet state. The 2009 flaring epoch is not as bright as the 1998 flare detected by \RXTE, which reached an X-ray flux of nearly 50 times higher than the quiet state (see Table~\ref{tab:xray-pow}). 

In 2011-2012 ($\beta$) and 2020 ($\gamma$) \pks was in a quiet state. It is worth noting that both data sets are perfectly compatible in flux and spectral shape, i.e., the data suggest a stable period of quiescence at similar flux levels and physical conditions in the jet. We discuss the X-ray data in detail in Sect.~\ref{sec-xray}

\begin{figure}[bth]
    \centering
    \includegraphics[width=\textwidth]{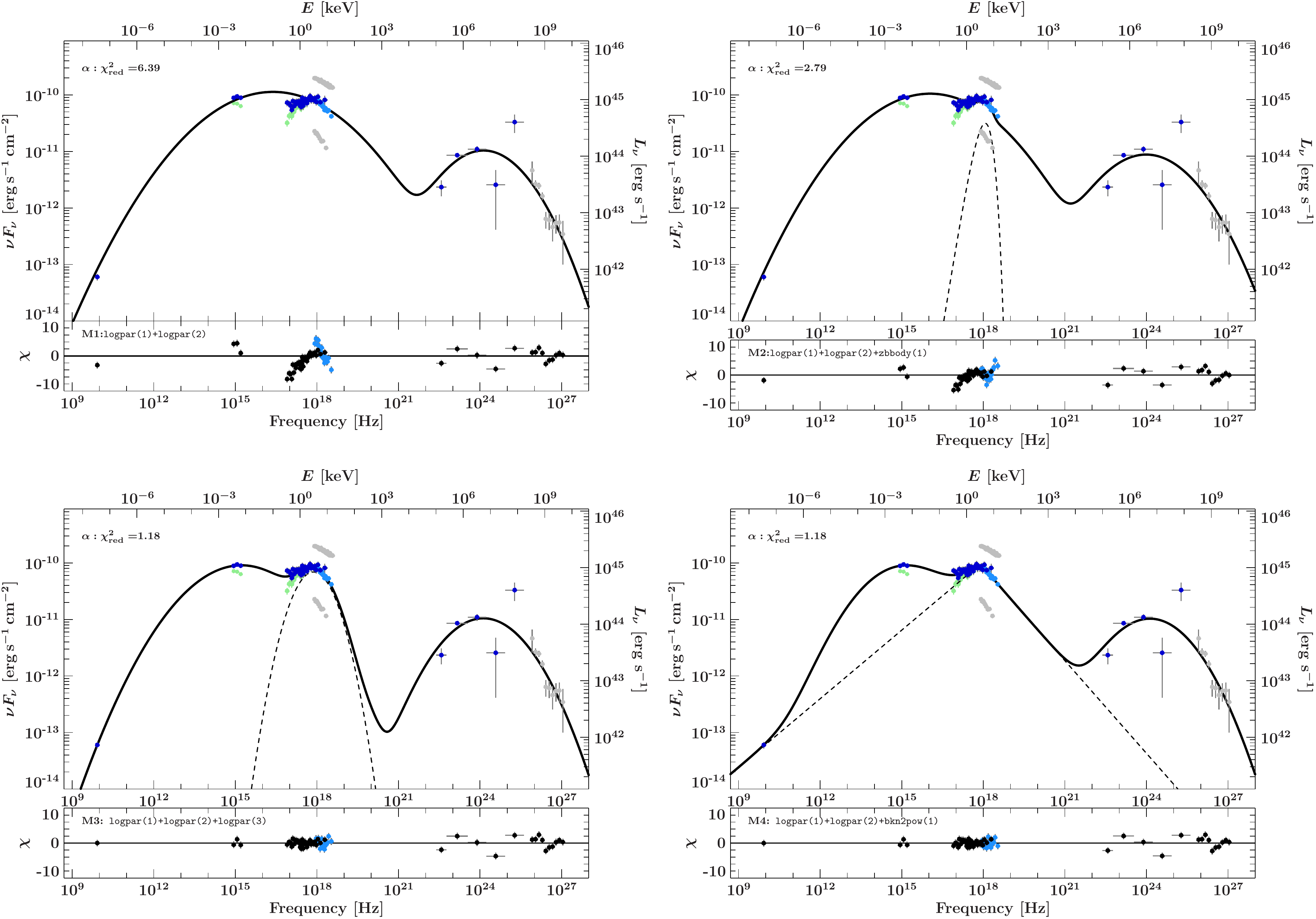}
    \caption{The broadband SED of \pks during the $\alpha$ (2008--2009) epoch is shown. Archival H.E.S.S. data (not contemporaneous) are shown in gray. 
    Multiple models and their residuals are given in each of the four panels and are shown with their associated residual panel below each SED.
    A simple logarithmic parabola (top left panel) or an added blackbody are not able to model the strong X-ray flux with its break.
    Both a third parabola (bottom left) or a broken powerlaw component (bottom right panel) are able to explain the data.
    }
    \label{fig:alpha-sed}
\end{figure}

The SEDs in the two quiet states ($\beta$ and $\gamma$) are very well described with logarithmic parabolas (see Table~\ref{tab:fit-results} for a complete list of fit results).
The best-fit empirical parabola models for each epoch are shown in Fig.~\ref{fig:full-sed}. The residuals are shown in the bottom panel, while the reduced $\chi^2$ values are also listed in the left side. We further examine the high-flux state ($\alpha$) of the source in Fig.~\ref{fig:alpha-sed} with the fit results provided in Table~\ref{tab:fit-resultsalpha}.
In all plots, the $\alpha$ data are shown in light blue, and the \RXTE data are shown in dark blue in order to distinguish between the two data sets.
The low-state data ($\beta$ and $\gamma$) are shown in dark green and turquoise, respectively. Archival data are shown in orange, while the reddened and absorbed data are shown in light gray. Error bars of all data sets are shown in dark gray.
Instrumental energy ranges are shown at the top of the plot with the position of the line along the vertical axis corresponding to the position of the labels (i.e., higher lines correspond to higher labels).

Fig.~\ref{fig:full-sed} shows the best-fit empirical logarithmic parabola models. In the $\beta$ state, a host galaxy blackbody has been added in order to describe the optical data.
The quiescent states are reasonably well described by the empirical model, including the newer 2020 $\gamma$ epoch.
It is worth noting that both the $\beta$ and the $\gamma$ data are compatible in flux and spectral shape. As the data were taken $\sim$ eight years apart, it suggests a stable mode of quiescence.
We confirm that the high state ($\alpha$) is incompatible with a simple empirical model. The X-ray band shows a peculiar, narrowly curved shape that is incompatible with the expected wide synchrotron-peak behavior. As far as we are aware, this is the only source to exhibit this peculiar feature in the X-ray band, as well as the extreme X-ray variability not accompanied by strong variability in other wavebands.
We investigate the broadband behavior of \pks in the $\alpha$ state in more detail in Fig.~\ref{fig:alpha-sed}, where we model the flaring state with a variety of empirical models in order to constrain the shape of the feature. We discuss the X-ray band in detail in Sect.~\ref{sec-xray} and possible origins of the peculiar feature in Sect.~\ref{sec:ori}.
Data used for fitting include the quasi-simultaneous data for $\alpha$ in addition to the archival H.E.S.S. data to constrain the high-energy peak. The H.E.S.S. data can be considered quasi-simultaneous with the $\alpha$ state.
The first model is identical to the empirical SED model shown in Fig.~\ref{fig:full-sed} and is not able to explain the shape of the X-ray spectrum.
The second model (M2, residuals shown in the third panel from the top) adds a blackbody model in an attempt to describe the narrowly curved feature with a thermal component. The blackbody itself is also shown as a dashed line. It is not able to adequately describe the shape as indicated by the residuals. The third model (M3, residuals shown in the fourth panel from the top) adds a third parabola to the two describing the synchrotron and high-energy peaks in order to describe the X-ray data. As it is most flexible in terms of how narrow the curvature is, it is able to describe the feature well as indicated by the residuals and reduced $\chi^2$ values. The fourth model (M4, bottom panel) adds a three-segment broken powerlaw, which is also able to explain the X-ray data well. However, it requires five parameters that are free to vary. This is likely too many, as indicated by the reduced $\chi^2$ value below 1 for all three datasets.
The best-fit values of the logarithmic parabola model are given in Table~\ref{tab:fit-resultsalpha}. Additional model features, such as the broken powerlaw, are described in Table~\ref{tab:fit-resultsalpha2}.

\begin{sidewaystable}\small
\centering
    \caption{Best-fit results for the logarithmic parabola fits to the various epochs of the \pks SED shown in Fig.~\ref{fig:full-sed}. If no uncertainty is given, the parameter was fixed to the given value. The absorbing column for the reddening and absorption model was frozen to the Galactic value of $N_\mathrm{H}=3.93\times10^{20}\,\mathrm{cm}^{-2}$. A blackbody fit was only used to describe a host galaxy signature in the $\beta$ state. The blackbody temperature has been corrected for the redshift of the source. The $E_p$ of the synchrotron peak is 0.05\,keV in all models. The uncertainties are statistical only.}    
    \begin{tabular}{l|lll|llllcccccl}
          & $N_\mathrm{sync}$ & $a_\mathrm{sync}$ & $b_\mathrm{sync}$ &  $N_\mathrm{HE}$ & $a_\mathrm{HE}$ & $b_\mathrm{HE}$ & $E_\mathrm{p,HE}$ & $N_\mathrm{blackbody}$ & $T_\mathrm{blackbody}$ & $\chi^2$ & $\chi^2_r$ & $F_{\mathrm{bol}}$\\
               &  &  &  & [keV] & [$10^{-10}$]  & &  & [$10^{-4}$] & [$10^{-3}$keV]&&& [erg\,s$^{-1}$\,cm$^{-2}$]\\
         \hline
         $\alpha$ & $27.7\pm0.7$ & $1.954\pm0.005$ & $0.0751^{+0.0018}_{-0.0017}$  & $0.24^{+0.08}_{-0.07}$ & $0.503^{+0.025}_{-0.026}$ & $0.185$ & 500 & -- & -- & 460.18 & 6.39 & $\left(1.173\pm0.019\right)\times10^{-9}$\\
         $\beta$ & $8.1\pm0.4$ & $2.310\pm0.013$ & $0.1269^{+0.0030}_{-0.0029}$ & $1.7\pm0.4$ & $0.579^{+0.021}_{-0.019}$ & 0.15 & 100 & $3.2\pm0.7$ & $1.14^{+0.13}_{-0.10}$ & 86.73 & 2.80 & $\left(4.94\pm0.10\right)\times10^{-10}$\\
         $\gamma$ & $8.4\pm0.4$ & $2.306^{+0.008}_{-0.009}$ & $0.1231\pm0.0023$  & $0.088\pm0.01$ & 0.0582 & 0.197 & 100 & -- & -- & 146.95 & 2.77 & $\left(4.53\pm0.21\right)\times10^{-10}$\\
    \end{tabular}
    \label{tab:fit-results}
    \vspace*{1.5cm}
    \caption{Best-fit results for the logarithmic parabola fits to the various models of the $\alpha$ SED shown in Fig.~\ref{fig:alpha-sed}. If no uncertainty is given, the parameter was fixed to the given value. The absorbing column for the reddening and absorption model was frozen to the Galactic value of $N_\mathrm{H}=3.93\times10^{20}\,\mathrm{cm}^{-2}$. The $E_p$ of the synchrotron peak is 0.05\,keV in all sources. The uncertainties are statistical only.}
    
    \begin{tabular}{l|lll|llllccl}
    & $N_\mathrm{sync}$ & $a_\mathrm{sync}$ & $b_\mathrm{sync}$ &  $N_\mathrm{HE}$ & $a_\mathrm{HE}$ & $b_\mathrm{HE}$ & $E_\mathrm{p,HE}$& $\chi^2$ & $\chi^2_r$ & $F_{\mathrm{bol}}$\\
               &  &  &  &  [$10^{-10}$]  & & & [keV] & & & [erg\,s$^{-1}$\,cm$^{-2}$]\\
         \hline
    M1 & $27.7\pm0.7$ & $1.954\pm0.005$ & $0.0751^{+0.0018}_{-0.0017}$  & $0.24^{+0.08}_{-0.07}$ & $0.503^{+0.025}_{-0.026}$ & $0.185$ & 500 & 460.18 & 6.39 & $\left(1.173\pm0.019\right)\times10^{-9}$\\
    M2 & $26.2\pm0.7$ & $2.007\pm0.009$ & $0.0844^{+0.0026}_{-0.0024}$  & $3.6^{+1.3}_{-1.0}$ & $0.618\pm0.024$ & 0.15 & 100 & 195.34 & 2.79 & $\left(1.075\pm0.022\right)\times10^{-9}$\\
    M3 & $19^{+4}_{-5}$ & $2.19^{+0.13}_{-0.08}$ & $0.113^{+0.018}_{-0.011}$ & $0.25^{+0.08}_{-0.07}$ & $0.505\pm0.025$ & 0.185 & 500 & 81.62 & 1.18 & $\left(1.007\pm0.022\right)\times10^{-9}$ \\
    M4 & $14.1^{+2.9}_{-4.6}$ & $2.34^{+0.21}_{-0.20}$ & $0.17^{+0.04}_{-0.07}$ & $0.24^{+0.08}_{-0.07}$ & $0.501\pm0.026$ & 0.185 & 500 & 77.62 & 1.18  & $\left(1.01^{+0.08}_{-0.06}\right)\times10^{-9}$\\
    \end{tabular}
    \label{tab:fit-resultsalpha}
    \vspace*{1.5cm}
 \caption{Best-fit results for the additional components of the various models of the $\alpha$ SED shown in Fig.~\ref{fig:alpha-sed}. If no uncertainty is given, the parameter was fixed to the given value. The absorbing column for the reddening and absorption model was frozen to the Galactic value of $N_\mathrm{H}=3.93\times10^{20}\,\mathrm{cm}^{-2}$. The uncertainties are statistical only.}
    
    \begin{tabular}{l|cccccccccccc}
   & $N_\mathrm{bb}$ & $T$ & $N_\mathrm{lp3}$ &  $a_\mathrm{lp3}$ & $b_\mathrm{lp3}$ & $E_\mathrm{p,lp3}$ & $N_{\mathrm{bknpow}}$ &$\Gamma_1$ & $\Gamma_2$& $\Gamma_3$ & $E_\mathrm{break,1}$& $E_\mathrm{break,2}$\\
    & [$10^{-4}$] & [keV] & & & & [keV] & & & & & [keV] &[keV]\\
         \hline
         M1 &  -- & --& -- & -- & -- & -- & --&-- &-- & -- & -- & --\\
         M2 & $5.8\pm0.6$ & $1.33\pm0.07$ & -- & -- & -- & -- &-- & --&-- &-- & -- & --\\
         M3 & -- & -- & $0.059^{+0.035}_{-0.029}$ & $0.83^{+0.30}_{-0.46}$ & $0.74^{+0.23}_{-0.15}$ & 0.5 & --&-- & -- & -- & -- & --\\ 
         M4 & -- & -- &-- & --& --& -- & $0.037^{+0.011}_{-0.017}$ & $1.596^{+0.015}_{-0.494}$ & $2.09^{+0.23}_{-0.50}$  & $2.55^{+0.09}_{-0.08}$ & $2.2^{+0.7}_{-1.6}$ &  $4.5^{+1.3}_{-0.6}$\\
    \end{tabular}
    \label{tab:fit-resultsalpha2}
\end{sidewaystable}

\subsection{X-ray data}
\label{sec-xray}
\begin{figure}[thb]
    \centering
    \includegraphics[width=\textwidth]{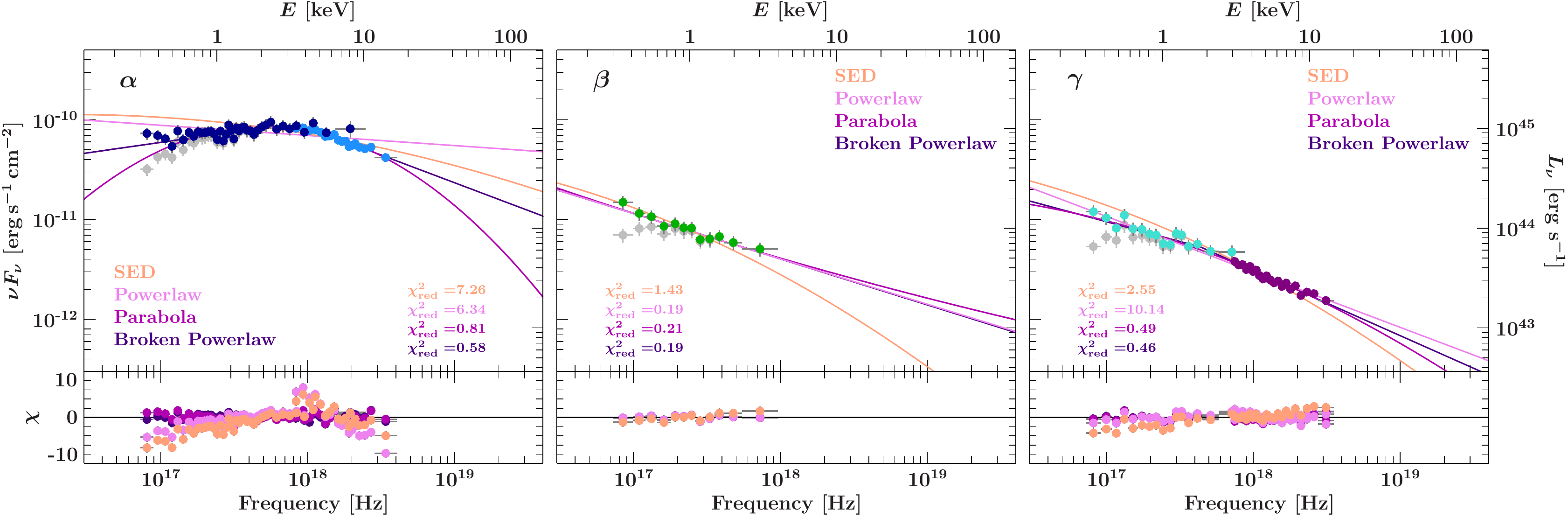}
    \caption{The X-ray spectrum of \pks. We include simultaneous X-ray data of \pks and the reddened and absorbed data in gray. The colors of the SED data correspond to the colors in the left panel of Fig.~\ref{fig:full-sed}. The three epochs are shown in the three panels. Four models are shown in orange, pink, magenta and purple in each of the panels with the corresponding reduced $\chi^2$ values. Residual panels are shown below the plots and are shown in the color of each model. The 0.5--10\,keV data in all three panels (dark blue, green, and turquoise) are \Swift/XRT data. Light blue data in the left panel are \RXTE/PCA data, while the purple data in the right panel are \nus data.}
    \label{fig:xspec}
\end{figure}

From the multiwavelength SED, it is clear that the measurements in most wavebands are relatively consistent in flux and shape between low and high-flux states.
A strong deviation from quiescence is only observed in the X-ray band.
In this section, we present additional modeling of the X-ray data, independent of the multiwavelength observations.
Fig.~\ref{fig:xspec} shows the X-ray band with the data of all three epochs in the same colors as in the SED. Various models are shown in shades of purple. All models include photoelectric absorption fixed to the Galactic value (see Tables~\ref{tab:xray-pow}, \ref{tab:xray-log}, and \ref{tab:xray-bknpower} for details).
The first model is identical to the SED model for comparative purposes. Secondly, we fit an absorbed powerlaw to all of the data. While it describes the quiescence data well for both states ($\beta$, $\gamma$), it is not able to adequately model the high-flux state $\alpha$.
The third model adds a third logarithmic parabola in order to describe the peculiar X-ray feature. It describes the X-ray data well as indicated in Fig.~\ref{fig:alpha-sed}. Despite also describing the other states well, $\chi^2_r<1$ indicates too many degrees of freedom in the model. The fourth model adds a three-segment broken powerlaw for the $\alpha$ state and a two-segment broken powerlaw for the other two states. All states are well described by the additional broken powerlaw, but the quiescence SEDs indicate a large number of degrees of freedom ($\chi^r_\mathrm{red}<1$), and the $\beta$ state find the same photon index for both sides, requiring one to be frozen due to the two parameters being correlated.
The models in Fig.~\ref{fig:xspec} are given in Tables~\ref{tab:xray-pow}, \ref{tab:xray-log}, and \ref{tab:xray-bknpower}.
In Table ~\ref{tab:xray-pow}, we show two powerlaw fits to the $\alpha$ state, one with a frozen absorbing column and one where $N_\mathrm{H}$ was left free to vary. A higher absorption is able to partially mimic the observed curved shape, as indicated by the improved $\chi^2$. From the residuals (not shown), it is clear that this is not a feasible model for the high flux state. Further, significant obscuration of a blazar jet is extremely unlikely.

\begin{table}[tbh]
    \centering
        \caption{Best-fit results of a powerlaw fit to the X-ray data of all epochs, shown in Fig.~\ref{fig:xspec}. Residuals can be found in the second residual panel. We include two fits to the $\alpha$ state. The second fit leaves the absorbing column free to vary in the fit. The uncertainties are statistical only.}
    \begin{tabular}{l|cccccl}
                Epoch &$N_\mathrm{H}$ & $N$ & $\Gamma_\mathrm{powerlaw}$ & $\chi$ & $\chi^2_{\mathrm{red}}$   & $F_{2-10\,\mathrm{keV}}$\\
                & [$10^{20}$\,cm$^{-2}$] &&&&& [erg s$^{-1}$ cm$^{-2}$]\\
                \hline
         $\alpha$ & 3.93 & $0.0503\pm0.0014$ & $2.102\pm0.017$ & 354.94 & 6.34 & $\left(1.099\pm0.016\right)\times10^{-10}$\\
         $\alpha$ (var. $N_\mathrm{H}$) & $17.5^{+2.5}_{-2.3}$ & $0.070\pm0.004$ & $2.286\pm0.030$ & 169.47 & 3.08 & $\left(1.170\pm0.019\right)\times10^{-10}$\\
         $\beta$ & 3.93 & $\left(4.8\pm0.4\right)\times10^{-3}$ & $2.45\pm0.14$ & 1.93 & 0.19 & $\left(6.3^{+1.4}_{-1.3}\right)\times10^{-12}$ \\
         $\gamma$ & 3.93 & $\left(4.11\pm0.23\right)\times10^{-3}$ & $2.56\pm0.04$ & 21.89 & 0.63 & $\left(4.68\pm0.14\right)\times10^{-12}$  \\
         1998 & 3.93 & $0.172\pm0.007$ & $2.258\pm0.023$ & 21.65 & 0.77 & $\left(2.99\pm0.04\right)\times10^{-10}$\\
    \end{tabular}
    \label{tab:xray-pow}
\end{table}
\begin{table}[tbh]
    \centering
        \caption{Best-fit results of a logarithmic parabola fit to the X-ray data of all epochs, shown in Fig.~\ref{fig:xspec}. Residuals can be found in the third residual panel. The absorbing column was frozen to the Galactic value of $N_\mathrm{H}=3.93\times10^{20}\,\mathrm{cm}^{-2}$.}
    \begin{tabular}{l|cccccccl}
        Epoch &$N_\mathrm{H}$ & $N$ & $a$ & $b$ & $E_p$ &$\chi^2$ & $\chi^2_\mathrm{red}$ & $F_{2-10\,\mathrm{keV}}$   \\
                & [$10^{20}$\,cm$^{-2}$] & & & & [keV] & & & [erg s$^{-1}$ cm$^{-2}$]\\
                \hline
         $\alpha$ & 3.93 & $0.0479^{+0.0018}_{-0.0019}$ & $1.68\pm0.06$ & $0.48\pm0.06$ & 1 & 44.73 & 0.81 & $\left(1.183\pm0.018\right)\times10^{-10}$\\
         $\beta$ & 3.93 & $\left(4.8^{+0.6}_{-0.5}\right)\times10^{-3}$ & $2.46\pm0.14$ & $0.0^{+0.6}_{-0.5}$ & 1 & 1.91 & 0.21 & $\left(6.4^{+3.4}_{-2.3}\right)\times10^{-12}$\\
         $\gamma$ & 3.93 & $\left(4.17^{+0.25}_{-0.24}\right)\times10^{-3}$ & $2.47\pm0.08$ & $0.12\pm0.09$ & 1 & 16.72 & 0.49 & $\left(4.80\pm0.17\right)\times10^{-12}$\\

    \end{tabular}
    \label{tab:xray-log}
\end{table}

\begin{table}[hbt]
    \centering
        \caption{Best-fit results of a three/two-segment broken powerlaw fit to the X-ray data of all epochs, shown in Fig.~\ref{fig:xspec}. Residuals can be found in the fourth residual panel. The uncertainties are statistical only.}
    \begin{tabular}{l|cccccclll}
    Epoch & $N$ & $\Gamma_1$ & $\Gamma_2$& $\Gamma_3$ & $E_\mathrm{break,1}$ & $E_\mathrm{break,2}$ &  $\chi^2$ & $\chi^2_\mathrm{red}$ & $F_{2-10\,\mathrm{keV}}$  \\
        & [$10^{-3}$\,cm$^{-2}$] & & & & [keV] & [kev] & &  & [erg s$^{-1}$ cm$^{-2}$]\\
        \hline
    $\alpha$ & $44.7^{+1.6}_{-1.9}$ & $1.79\pm0.09$ & $2.11^{+0.22}_{-0.15}$ & $2.56\pm0.07$ & $2.4^{+1.3}_{-1.1}$ & 4.5 & 30.69 & 0.58 & 
$\left(1.177\pm0.024\right)\times10^{-10}$\\
  $\beta$ & $4.8\pm 0.4$ & $2.46\pm0.15$ & 2.46 & -- & 2.5 & -- &  1.90 & 0.21 & $\left(6.3\pm1.0\right)\times10^{-12}$\\

  $\gamma$ & $4.2\pm0.4$ & $2.40^{+0.14}_{-0.17}$ & $2.653^{+0.089}_{-0.0002}$ & -- & $1.8^{+2.2}_{-0.9}$ & -- & 15.08 & 0.46 & $\left(4.86^{+0.21}_{-0.20}\right)\times10^{-12}$\\
    \end{tabular}
    \label{tab:xray-bknpower}
\end{table}

From the SED and the X-ray spectrum, it is noteworthy how consistent quiescence is in \pks. Between the 2011--2012 and 2020 epochs, the source either did not vary in flux, or returned to quiescence with identical fluxes and spectral indices. The 2020 data include our \nus data, which is consistent with \Swift/XRT data in flux and spectral shape of both low-flux epochs ($\beta$ and $\gamma$), but not consistent with the \Swift/BAT data.

\subsection{Possible explanations of the peculiar X-ray feature in the high-flux state}
\label{sec:ori}
X-ray observations of \pks reveal a peculiar feature in high-flux states, while at other wavelengths, the source varies little, and the spectral shape does not change significantly.
During high-flux states, the X-ray band is characterized by extreme flares with X-ray fluxes reaching 10--50 times greater than fluxes during quiescent periods.
We note that the host galaxy of \pks is a giant elliptical galaxy \citep{Scarpa2000}, which is unlikely to contribute to the X-ray emission through active star formation or other processes that would outshine the relativistic jet.
\clearpage
\subsubsection{Instrumental feature}
\Swift/XRT data have been examined and treated for pile-up, which cannot sufficiently explain the curvature and high-flux state. Furthermore, \RXTE data are consistent with \Swift/XRT, indicating that an instrumental artifact or calibration issues can be ruled out. Additional archival data, such as the EXOSAT data presented in \citet{Giommi1990}, show extreme flaring periods, strongly suggesting that this is not an instrumental issue.

\subsubsection{Thermal feature}
A single-temperature blackbody cannot adequately model the data (see Fig.~\ref{fig:xspec}, Table~\ref{tab:fit-resultsalpha2}), which indicates that the origin of the feature is likely nonthermal in origin. Temperatures are higher than possible for a host galaxy or an accretion disk \citep{Malkan1983}.

\subsubsection{Hadronic origin}
In principle, the X-ray emission could result from synchrotron emission by secondary pairs generated by hadronic cascades \citep[e.g.,][]{Boettcher2013,Keivani2018}. Nonetheless, this scenario requires extreme physical parameters for the observed spectrum here. The roughly comparable optical and X-ray flux indicate very powerful secondary synchrotron emission, which generally points to efficient photomeson production. 
However, the target photons cannot be the result of synchrotron radiation from the primary electrons located in the hadronic emission region. In such a scenario, the primary electrons would upscatter the X-rays to $\gamma$-rays, which would have been observable by \Fermi/LAT and TeV telescopes.
To avoid this, the hadronic emission region could be located in the broad line region whose flux may be lower than the primary electron synchrotron emission \citep[so-called masquerading BL Lacs; ][]{Giommi2013}. In this situation, the kinetic luminosity of protons must be very high to produce adequate secondary pairs to match the observed X-ray flux. This scenario would likely be disfavored due to the unrealistic jet powers necessitated by detailed spectral modeling. Additionally, the proton cooling due to photomeson processes is generally slow, which is inconsistent with the fast X-ray variability that were previously observed in this source \citep{Giommi1990,Sambruna1995}. If a hadronic model were plausible in a certain parameter space, then the X-ray flux in the high state would likely be accompanied by a comparably high neutrino flux. This secondary pair synchrotron component can be strongly polarized, which may be detected by X-ray polarimeters such as \textsl{IXPE} \citep[e.g.,][]{Zhang2017}.

\subsubsection{Additional jet emission component and magnetic reconnection}

Another possibility is that the X-ray spectrum originates from synchrotron emission by a population of high-energy nonthermal electrons in an environment with a high magnetic field. Since the ratio of Compton scattering flux over synchrotron is proportional to the ratio of photon energy density over magnetic energy density, a high magnetic field can suppress the inverse Compton scattering, resulting in negligible contribution to $\gamma$-rays \citep[see, e.g.,][]{Rybicki1985}. 
By comparing the X-ray flux and TeV $\gamma$-ray flux, we expect that the magnetic energy density in this orphan component should exceed $\gtrsim 100$ times the photon energy density. In contrast, the same ratio derived from the optical and {\Fermi} $\gamma$-ray fluxes is $\sim 10$. This suggests that the high state may result from the emergence of a new emission blob with very strong magnetic field, which is not likely to be located in the emission region responsible for the low state. 
Given the rather hard spectral shape in the X-ray component, these high-energy electrons are probably accelerated via magnetic reconnection. \citet{Zhang2021} have shown that reconnection in a highly magnetized environment can lead to very hard X-ray synchrotron spectra. 
Both the flux and polarization can be strongly variable in reconnection, which may be examined by {\textsl{IXPE}} and future X-ray polarimetry instruments. Alternatively, turbulence in a highly magnetized environment may also accelerate high-energy electrons. 
However, the electron spectrum is usually much softer than reconnection, which may be disfavored by spectral modeling \citep{Comisso2018}.

\section{Conclusions} \label{sec:conc}

We discuss the broadband SED of the southern BL Lac object \pks. While the SEDs during two epochs have already been shown by \citet{krauss2016}, we present new data taken in 2020, including the first hard X-ray spectrum with \nus data.
We find that quiescence is stable in the source, since the 2020 data are perfectly compatible with the 2012 low-state data.
The most peculiar feature of \pks is its extreme X-ray variability that is not accompanied by large flux changes in other wavebands and exhibits a peculiar spectrum.
The SED in the high state cannot be explained with a typical logarithmic parabola that, though empirical, is able to describe blazar spectra well \citep{krauss2016}.
Currently, the most likely explanation seems to be an additional jet component that is not co-spatial with the low-state emission region. The X-ray flaring activity can be explained through magnetic reconnection, which also predicts strong X-ray polarization variability.
In order to understand the flaring in \pks, additional data sets during the high state are crucial.
\nus and X-ray polarimetry data in the high state and data in additional wavebands (e.g., MeV) would allow us to better constrain the physical mechanisms at play.
As the extreme source variability is only present in the X-ray band and does not exhibit frequent flaring, it is difficult to monitor its flux changes and detect a bright state due to the lack of an all-sky monitor at X-ray energies capable of providing frequent flux measurements.

\section{Acknowledgements}
We thank Tonia Venters and Matthew Kerr for their helpful comments that improved the manuscript.
We thank Victoria Grinberg for discussions and comments.
O.C. is supported by the NASA program NNH19ZDA001N-NUSTAR through grant \#19-NUSTAR19-0040.
H.Z. is supported by an appointment to the NASA Postdoctoral Program at Goddard Space Flight Center, administered by Oak Ridge Associated Universities under contract with NASA.
This research has made use of a collection of ISIS functions (ISISscripts) provided by ECAP/Remeis observatory and MIT (http://www.sternwarte.uni-erlangen.de/isis/).
Part of this work is based on archival data, software or online services provided by the Space Science Data Center -- ASI.

The \textsl{Fermi}/LAT Collaboration  acknowledges generous ongoing support from a
number of agencies and institutes that have supported both the development and the operation of the LAT as well as scientific data analysis.
These include the National Aeronautics and Space Administration and the Department of Energy in the United States, the Commissariat \`a l'Energie Atomique and the Centre National de la Recherche Scientifique / Institut National de Physique Nucl\'eaire et de Physique des Particules in France, the Agenzia Spaziale Italiana and the
Istituto Nazionale di Fisica Nucleare in Italy, the Ministry of Education, Culture, Sports, Science and Technology (MEXT), High Energy Accelerator Research Organization (KEK) and Japan Aerospace Exploration Agency (JAXA) in Japan, and the K. A. Wallenberg Foundation, the Swedish Research Council and the Swedish National Space Board in Sweden.
Additional support for science analysis during the operations phase is gratefully acknowledged from the Istituto Nazionale di Astrofisica in Italy and the Centre National d'Etudes Spatiales in France. This work performed in part under DOE Contract DE- AC02-76SF00515.

\bibliographystyle{jwaabib}
\bibliography{mnemonic,aa_abbrv,all}

\begin{thebibliography}{}
\expandafter\ifx\csname natexlab\endcsname\relax\def\natexlab#1{#1}\fi
\providecommand{\url}[1]{\href{#1}{#1}}
\providecommand{\dodoi}[1]{doi:~\href{http://doi.org/#1}{\nolinkurl{#1}}}
\providecommand{\doeprint}[1]{\href{http://ascl.net/#1}{\nolinkurl{http://ascl.net/#1}}}
\providecommand{\doarXiv}[1]{\href{https://arxiv.org/abs/#1}{\nolinkurl{https://arxiv.org/abs/#1}}}

\bibitem[{{Abdollahi} {et~al.}(2020){Abdollahi}, {Acero}, {Ackermann},
  {Ajello}, {Atwood}, {Axelsson}, {Baldini}, {Ballet}, {Barbiellini},
  {Bastieri}, {Becerra Gonzalez}, {Bellazzini}, {Berretta}, {Bissaldi},
  {Blandford}, {Bloom}, {Bonino}, {Bottacini}, {Brandt}, {Bregeon}, {Bruel},
  {Buehler}, {Burnett}, {Buson}, {Cameron}, {Caputo}, {Caraveo}, {Casandjian},
  {Castro}, {Cavazzuti}, {Charles}, {Chaty}, {Chen}, {Cheung}, {Chiaro},
  {Ciprini}, {Cohen-Tanugi}, {Cominsky}, {Coronado-Bl{\'a}zquez}, {Costantin},
  {Cuoco}, {Cutini}, {D'Ammando}, {DeKlotz}, {de la Torre Luque}, {de Palma},
  {Desai}, {Digel}, {Di Lalla}, {Di Mauro}, {Di Venere}, {Dom{\'\i}nguez},
  {Dumora}, {Fana Dirirsa}, {Fegan}, {Ferrara}, {Franckowiak}, {Fukazawa},
  {Funk}, {Fusco}, {Gargano}, {Gasparrini}, {Giglietto}, {Giommi}, {Giordano},
  {Giroletti}, {Glanzman}, {Green}, {Grenier}, {Griffin}, {Grondin}, {Grove},
  {Guiriec}, {Harding}, {Hayashi}, {Hays}, {Hewitt}, {Horan},
  {J{\'o}hannesson}, {Johnson}, {Kamae}, {Kerr}, {Kocevski}, {Kovac'evic'},
  {Kuss}, {Landriu}, {Larsson}, {Latronico}, {Lemoine-Goumard}, {Li},
  {Liodakis}, {Longo}, {Loparco}, {Lott}, {Lovellette}, {Lubrano}, {Madejski},
  {Maldera}, {Malyshev}, {Manfreda}, {Marchesini}, {Marcotulli},
  {Mart{\'\i}-Devesa}, {Martin}, {Massaro}, {Mazziotta}, {McEnery}, {Mereu},
  {Meyer}, {Michelson}, {Mirabal}, {Mizuno}, {Monzani}, {Morselli},
  {Moskalenko}, {Negro}, {Nuss}, {Ojha}, {Omodei}, {Orienti}, {Orlando},
  {Ormes}, {Palatiello}, {Paliya}, {Paneque}, {Pei}, {Pe{\~n}a-Herazo},
  {Perkins}, {Persic}, {Pesce-Rollins}, {Petrosian}, {Petrov}, {Piron}, {Poon},
  {Porter}, {Principe}, {Rain{\`o}}, {Rando}, {Razzano}, {Razzaque}, {Reimer},
  {Reimer}, {Remy}, {Reposeur}, {Romani}, {Saz Parkinson}, {Schinzel},
  {Serini}, {Sgr{\`o}}, {Siskind}, {Smith}, {Spandre}, {Spinelli}, {Strong},
  {Suson}, {Tajima}, {Takahashi}, {Tak}, {Thayer}, {Thompson}, {Tibaldo},
  {Torres}, {Torresi}, {Valverde}, {Van Klaveren}, {van Zyl}, {Wood},
  {Yassine}, \& {Zaharijas}}]{fermi_4fgl}
{Abdollahi}, S., {Acero}, F., {Ackermann}, M., {et~al.} 2020, \apjs, 247, 33,
  \dodoi{10.3847/1538-4365/ab6bcb}

\bibitem[{{Atwood} {et~al.}(2009){Atwood}, {Abdo}, {Ackermann}, {Althouse},
  {Anderson}, {Axelsson}, {Baldini}, {Ballet}, {Band}, {Barbiellini},
  {Bartelt}, {Bastieri}, {Baughman}, {Bechtol}, {B{\'e}d{\'e}r{\`e}de},
  {Bellardi}, {Bellazzini}, {Berenji}, {Bignami}, {Bisello}, {Bissaldi},
  {Blandford}, {Bloom}, {Bogart}, {Bonamente}, {Bonnell}, {Borgland},
  {Bouvier}, {Bregeon}, {Brez}, {Brigida}, {Bruel}, {Burnett}, {Busetto},
  {Caliandro}, {Cameron}, {Caraveo}, {Carius}, {Carlson}, {Casandjian},
  {Cavazzuti}, {Ceccanti}, {Cecchi}, {Charles}, {Chekhtman}, {Cheung},
  {Chiang}, {Chipaux}, {Cillis}, {Ciprini}, {Claus}, {Cohen-Tanugi},
  {Condamoor}, {Conrad}, {Corbet}, {Corucci}, {Costamante}, {Cutini}, {Davis},
  {Decotigny}, {DeKlotz}, {Dermer}, {de Angelis}, {Digel}, {do Couto e Silva},
  {Drell}, {Dubois}, {Dumora}, {Edmonds}, {Fabiani}, {Farnier}, {Favuzzi},
  {Flath}, {Fleury}, {Focke}, {Funk}, {Fusco}, {Gargano}, {Gasparrini},
  {Gehrels}, {Gentit}, {Germani}, {Giebels}, {Giglietto}, {Giommi}, {Giordano},
  {Glanzman}, {Godfrey}, {Grenier}, {Grondin}, {Grove}, {Guillemot}, {Guiriec},
  {Haller}, {Harding}, {Hart}, {Hays}, {Healey}, {Hirayama}, {Hjalmarsdotter},
  {Horn}, {Hughes}, {J{\'o}hannesson}, {Johansson}, {Johnson}, {Johnson},
  {Johnson}, {Johnson}, {Kamae}, {Katagiri}, {Kataoka}, {Kavelaars}, {Kawai},
  {Kelly}, {Kerr}, {Klamra}, {Kn{\"o}dlseder}, {Kocian}, {Komin}, {Kuehn},
  {Kuss}, {Landriu}, {Latronico}, {Lee}, {Lee}, {Lemoine-Goumard}, {Lionetto},
  {Longo}, {Loparco}, {Lott}, {Lovellette}, {Lubrano}, {Madejski}, {Makeev},
  {Marangelli}, {Massai}, {Mazziotta}, {McEnery}, {Menon}, {Meurer},
  {Michelson}, {Minuti}, {Mirizzi}, {Mitthumsiri}, {Mizuno}, {Moiseev},
  {Monte}, {Monzani}, {Moretti}, {Morselli}, {Moskalenko}, {Murgia},
  {Nakamori}, {Nishino}, {Nolan}, {Norris}, {Nuss}, {Ohno}, {Ohsugi}, {Omodei},
  {Orlando}, {Ormes}, {Paccagnella}, {Paneque}, {Panetta}, {Parent}, {Pearce},
  {Pepe}, {Perazzo}, {Pesce-Rollins}, {Picozza}, {Pieri}, {Pinchera}, {Piron},
  {Porter}, {Poupard}, {Rain{\`o}}, {Rando}, {Rapposelli}, {Razzano}, {Reimer},
  {Reimer}, {Reposeur}, {Reyes}, {Ritz}, {Rochester}, {Rodriguez}, {Romani},
  {Roth}, {Russell}, {Ryde}, {Sabatini}, {Sadrozinski}, {Sanchez}, {Sander},
  {Sapozhnikov}, {Parkinson}, {Scargle}, {Schalk}, {Scolieri}, {Sgr{\`o}},
  {Share}, {Shaw}, {Shimokawabe}, {Shrader}, {Sierpowska-Bartosik}, {Siskind},
  {Smith}, {Smith}, {Spandre}, {Spinelli}, {Starck}, {Stephens}, {Strickman},
  {Strong}, {Suson}, {Tajima}, {Takahashi}, {Takahashi}, {Tanaka}, {Tenze},
  {Tether}, {Thayer}, {Thayer}, {Thompson}, {Tibaldo}, {Tibolla}, {Torres},
  {Tosti}, {Tramacere}, {Turri}, {Usher}, {Vilchez}, {Vitale}, {Wang},
  {Watters}, {Winer}, {Wood}, {Ylinen}, \& {Ziegler}}]{fermi2009}
{Atwood}, W.~B., {Abdo}, A.~A., {Ackermann}, M., {et~al.} 2009, \apj, 697,
  1071, \dodoi{10.1088/0004-637X/697/2/1071}

\bibitem[{{Baumgartner} {et~al.}(2013){Baumgartner}, {Tueller}, {Markwardt},
  {Skinner}, {Barthelmy}, {Mushotzky}, {Evans}, \& {Gehrels}}]{bat}
{Baumgartner}, W.~H., {Tueller}, J., {Markwardt}, C.~B., {et~al.} 2013, apjs,
  207, 19

\bibitem[{{Beringer} {et~al.}(2012){Beringer}, {Arguin}, {Barnett}, {Copic},
  {Dahl}, {Groom}, {Lin}, {Lys}, {Murayama}, {Wohl}, \& {Yao}}]{Beringer2012}
{Beringer}, J., {Arguin}, J.~F., {Barnett}, R.~M., {et~al.} 2012, prd, 86,
  010001

\bibitem[{Blandford {et~al.}(1978)Blandford, Rees, \& Wolfe}]{blandford1978}
Blandford, R., Rees, M., \& Wolfe, A. 1978, AM Wolfe, Ed, 328

\bibitem[{{B{\"o}ttcher} {et~al.}(2013){B{\"o}ttcher}, {Reimer}, {Sweeney}, \&
  {Prakash}}]{Boettcher2013}
{B{\"o}ttcher}, M., {Reimer}, A., {Sweeney}, K., \& {Prakash}, A. 2013, \apj,
  768, 54, \dodoi{10.1088/0004-637X/768/1/54}

\bibitem[{{Comisso} \& {Sironi}(2018)}]{Comisso2018}
{Comisso}, L., \& {Sironi}, L. 2018, \prl, 121, 255101,
  \dodoi{10.1103/PhysRevLett.121.255101}

\bibitem[{{Gaia Collaboration}(2018)}]{gaia_dr2}
{Gaia Collaboration}. 2018, VizieR Online Data Catalog, I/345

\bibitem[{{Gardenier} \& {Uttley}(2018)}]{Gardenier18}
{Gardenier}, D.~W., \& {Uttley}, P. 2018, \mnras, 481, 3761,
  \dodoi{10.1093/mnras/sty2524}

\bibitem[{{Gehrels} {et~al.}(2004){Gehrels}, {Chincarini}, {Giommi}, {Mason},
  {Nousek}, {Wells}, {White}, {Barthelmy}, {Burrows}, {Cominsky}, {Hurley},
  {Marshall}, {M{\'e}sz{\'a}ros}, {Roming}, {Angelini}, {Barbier}, {Belloni},
  {Campana}, {Caraveo}, {Chester}, {Citterio}, {Cline}, {Cropper}, {Cummings},
  {Dean}, {Feigelson}, {Fenimore}, {Frail}, {Fruchter}, {Garmire}, {Gendreau},
  {Ghisellini}, {Greiner}, {Hill}, {Hunsberger}, {Krimm}, {Kulkarni}, {Kumar},
  {Lebrun}, {Lloyd-Ronning}, {Markwardt}, {Mattson}, {Mushotzky}, {Norris},
  {Osborne}, {Paczynski}, {Palmer}, {Park}, {Parsons}, {Paul}, {Rees},
  {Reynolds}, {Rhoads}, {Sasseen}, {Schaefer}, {Short}, {Smale}, {Smith},
  {Stella}, {Tagliaferri}, {Takahashi}, {Tashiro}, {Townsley}, {Tueller},
  {Turner}, {Vietri}, {Voges}, {Ward}, {Willingale}, {Zerbi}, \&
  {Zhang}}]{swift}
{Gehrels}, N., {Chincarini}, G., {Giommi}, P., {et~al.} 2004, apj, 611, 1005

\bibitem[{{Giommi} {et~al.}(1990){Giommi}, {Barr}, {Garilli}, {Maccagni}, \&
  {Pollock}}]{Giommi1990}
{Giommi}, P., {Barr}, P., {Garilli}, B., {Maccagni}, D., \& {Pollock}, A.~M.~T.
  1990, apj, 356, 432

\bibitem[{{Giommi} {et~al.}(2013){Giommi}, {Padovani}, \&
  {Polenta}}]{Giommi2013}
{Giommi}, P., {Padovani}, P., \& {Polenta}, G. 2013, mnras, 431, 1914

\bibitem[{{H.~E.~S.~S. Collaboration} {et~al.}(2010){H.~E.~S.~S.
  Collaboration}, {Acero}, {Aharonian}, {Akhperjanian}, {Anton}, {Barres de
  Almeida}, \& {Bazer-Bachi}}]{hess10}
{H.~E.~S.~S. Collaboration}, {Acero}, F., {Aharonian}, F., {et~al.} 2010, aap,
  511, A52

\bibitem[{{H.~E.~S.~S. Collaboration} \& {Fermi-LAT
  Collaboration}(2011)}]{hess2011}
{H.~E.~S.~S. Collaboration}, \& {Fermi-LAT Collaboration}. 2011, aap, 533, A110

\bibitem[{{Harrison} {et~al.}(2013){Harrison}, {Craig}, \&
  {Christensen}}]{nustar}
{Harrison}, F.~A., {Craig}, W.~W., \& {Christensen}, F.~E. 2013, ApJ, 770, 103

\bibitem[{{HI4PI Collaboration} {et~al.}(2016){HI4PI Collaboration}, {Ben
  Bekhti}, {Fl{\"o}er}, {Keller}, {Kerp}, {Lenz}, {Winkel}, {Bailin},
  {Calabretta}, {Dedes}, {Ford}, {Gibson}, {Haud}, {Janowiecki}, {Kalberla},
  {Lockman}, {McClure-Griffiths}, {Murphy}, {Nakanishi}, {Pisano}, \&
  {Staveley-Smith}}]{hi4pi}
{HI4PI Collaboration}, {Ben Bekhti}, N., {Fl{\"o}er}, L., {et~al.} 2016, aap,
  594, A116

\bibitem[{{Houck} \& {Denicola}(2000)}]{Houck2000}
{Houck}, J.~C., \& {Denicola}, L.~A. 2000, in Astronomical Society of the
  Pacific Conference Series, Vol. 216, Astronomical Data Analysis Software and
  Systems IX, ed. N.~{Manset}, C.~{Veillet}, \& D.~{Crabtree}, 591

\bibitem[{{Keeney} {et~al.}(2018){Keeney}, {Stocke}, {Pratt}, {Davis},
  {Syphers}, {Danforth}, {Shull}, {Froning}, {Green}, {Penton}, \&
  {Savage}}]{keeny2018}
{Keeney}, B.~A., {Stocke}, J.~T., {Pratt}, C.~T., {et~al.} 2018, \apjs, 237, 11

\bibitem[{{Keivani} {et~al.}(2018){Keivani}, {Murase}, {Petropoulou}, {Fox},
  {Cenko}, {Chaty}, {Coleiro}, {DeLaunay}, {Dimitrakoudis}, {Evans}, {Kennea},
  {Marshall}, {Mastichiadis}, {Osborne}, {Santander}, {Tohuvavohu}, \&
  {Turley}}]{Keivani2018}
{Keivani}, A., {Murase}, K., {Petropoulou}, M., {et~al.} 2018, \apj, 864, 84,
  \dodoi{10.3847/1538-4357/aad59a}

\bibitem[{{Krau{\ss}} {et~al.}(2016){Krau{\ss}}, {Wilms}, {Kadler}, \&
  {Ojha}}]{krauss2016}
{Krau{\ss}}, F., {Wilms}, J., {Kadler}, M., \& {Ojha}, R. 2016, A\&A, 591

\bibitem[{{Lin} {et~al.}(1999){Lin}, {Bertsch}, {Bloom}, {Esposito}, {Hartman},
  {Hunter}, {Kanbach}, {Kniffen}, {Mayer-Hasselwander}, {Michelson},
  {Mukherjee}, {M{\"u}cke}, {Nolan}, {Pohl}, {Reimer}, {Schneid}, {Thompson},
  \& {Tompkins}}]{egret}
{Lin}, Y.~C., {Bertsch}, D.~L., {Bloom}, S.~D., {et~al.} 1999, apj, 525, 191

\bibitem[{{Malkan}(1983)}]{Malkan1983}
{Malkan}, M.~A. 1983, ApJ, 268, 582

\bibitem[{{Mattox} {et~al.}(1996){Mattox}, {Bertsch}, {Chiang}, {Dingus},
  {Digel}, {Esposito}, {Fierro}, \& {Hartman}}]{mattox1996}
{Mattox}, J.~R., {Bertsch}, D.~L., {Chiang}, J., {et~al.} 1996, \apj, 461, 396,
  \dodoi{10.1086/177068}

\bibitem[{{Nowak} {et~al.}(2005){Nowak}, {Wilms}, {Heinz}, {Pooley},
  {Pottschmidt}, \& {Corbel}}]{Nowak2005}
{Nowak}, M.~A., {Wilms}, J., {Heinz}, S., {et~al.} 2005, apj, 626, 1006

\bibitem[{{Nowak} {et~al.}(2012){Nowak}, {Neilsen}, {Markoff}, {Baganoff},
  {Porquet}, {Grosso}, {Levin}, {Houck}, {Eckart}, {Falcke}, {Ji}, {Miller}, \&
  {Wang}}]{Nowak2012}
{Nowak}, M.~A., {Neilsen}, J., {Markoff}, S.~B., {et~al.} 2012, apj, 759, 95

\bibitem[{{Perlman} {et~al.}(1999){Perlman}, {Madejski}, {Stocke}, \&
  {Rector}}]{Perlman1999}
{Perlman}, E.~S., {Madejski}, G., {Stocke}, J.~T., \& {Rector}, T.~A. 1999,
  apjl, 523, L11

\bibitem[{{Predehl} \& {Schmitt}(1995)}]{Predehl1995}
{Predehl}, P., \& {Schmitt}, J.~H.~M.~M. 1995, aap, 293, 889

\bibitem[{{Rybicki} \& {Lightman}(1985)}]{Rybicki1985}
{Rybicki}, G.~B., \& {Lightman}, A.~P. 1985, {Radiative processes in
  astrophysics.} (John Wiley \& Sons)

\bibitem[{{Sambruna} {et~al.}(1995){Sambruna}, {Urry}, {Ghisellini}, \&
  {Maraschi}}]{Sambruna1995}
{Sambruna}, R.~M., {Urry}, C.~M., {Ghisellini}, G., \& {Maraschi}, L. 1995,
  apj, 449, 567

\bibitem[{{Scarpa} {et~al.}(2000){Scarpa}, {Urry}, {Padovani}, {Calzetti}, \&
  {O'Dowd}}]{Scarpa2000}
{Scarpa}, R., {Urry}, C.~M., {Padovani}, P., {Calzetti}, D., \& {O'Dowd}, M.
  2000, ApJ, 544, 258

\bibitem[{{Tagliaferri} {et~al.}(2001){Tagliaferri}, {Ghisellini}, {Giommi},
  {Celotti}, {Chiaberge}, {Chiappetti}, {Glass}, {Maraschi}, {Tavecchio},
  {Treves}, \& {Wolter}}]{Tagliaferri2001}
{Tagliaferri}, G., {Ghisellini}, G., {Giommi}, P., {et~al.} 2001, aap, 368, 38

\bibitem[{{Urry} \& {Padovani}(1995)}]{urry1995}
{Urry}, C.~M., \& {Padovani}, P. 1995, \pasp, 107, 803, \dodoi{10.1086/133630}

\bibitem[{{Verner} {et~al.}(1996){Verner}, {Ferland}, {Korista}, \&
  {Yakovlev}}]{Verner1996}
{Verner}, D.~A., {Ferland}, G.~J., {Korista}, K.~T., \& {Yakovlev}, D.~G. 1996,
  apj, 465, 487

\bibitem[{{Wakely} \& {Horan}(2008)}]{tevcat}
{Wakely}, S.~P., \& {Horan}, D. 2008, in International Cosmic Ray Conference,
  Vol.~3, International Cosmic Ray Conference, 1341--1344

\bibitem[{{Wall} {et~al.}(1975){Wall}, {Shimmins}, \& {Bolton}}]{parkes}
{Wall}, J.~V., {Shimmins}, A.~J., \& {Bolton}, J.~G. 1975, Australian Journal
  of Physics Astrophysical Supplement, 34, 55

\bibitem[{{Wilms} {et~al.}(2000){Wilms}, {Allen}, \& {McCray}}]{Wilms2000}
{Wilms}, J., {Allen}, A., \& {McCray}, R. 2000, apj, 542, 914

\bibitem[{{Zhang}(2017)}]{Zhang2017}
{Zhang}, H. 2017, Galaxies, 5, 32, \dodoi{10.3390/galaxies5030032}

\bibitem[{{Zhang} {et~al.}(2021){Zhang}, {Li}, {Giannios}, \&
  {Guo}}]{Zhang2021}
{Zhang}, H., {Li}, X., {Giannios}, D., \& {Guo}, F. 2021, apj, 912, 129

\end{thebibliography}

\end{CJK*}
\end{document}